\documentclass{aastex}
\usepackage{emulateapj5}

\newcommand {\ergs} {erg s$^{-1}$}
\newcommand {\ergcms} {erg cm$^{-2}$ s$^{-1}$}
\slugcomment{Accepted for publication in ApJL}
\begin{document}
\title{The Consequences of the Cosmic Star-Formation Rate: X-ray Number Counts}
 
\author{A. Ptak\altaffilmark{1}, R. Griffiths}
\affil{Carnegie Mellon University, Dept. of Physics, Pittsburgh, PA 15213}
\author{N. White}
\affil{NASA/GSFC LHEA, Greenbelt, MD 20771}
\author{P. Ghosh}
\affil{Tata Institute of Fundamental Research, Bombay 400 005, India}
\altaffiltext{1}{Present address: The Johns Hopkins University,
Dept. of Physics and Astronomy, 3400 N. Charles St., Baltimore, MD
21218; ptak@pha.jhu.edu}
\begin{abstract}
We discuss the observable consequences for the detection of galaxies 
in the X-ray bandpass resulting from
a peak in the cosmic star-formation rate at a redshift $>$ 1.  Following White
\& Ghosh, we assume a large evolution in the X-ray/B luminosity ratio at
$z \sim 0.5-1.5$
resulting from the X-ray binaries that have evolved from stars formed at $z
> 1-2$. Using the HDF-N redshift survey data and the locally observed
X-ray/B luminosity ratio as a guide, we estimate a median X-ray flux (2-10
keV) on the order of $8 \times 10^{-18}$ \ergcms ~for galaxies in the HDF-N,
which is consistent with a
signal derived from a stacking analysis of the HDF-N Chandra data by Brandt
et al. (2001).  We also predict the number counts in deep X-ray surveys
expected from normal galaxies at high redshift.
\end{abstract}
\section{Introduction}

The X-ray emission of nearby normal spiral
galaxies has been observed to be dominated
by X-ray binaries (XRB; see Fabbiano 1989 for a
review).  There is a strong correlation between the optical and X-ray
luminosities of galaxies which is also consistent with an XRB population
dominating the X-ray emission (i.e., assuming that XRB constitute an ~ constant fraction of stars in a galaxy).   Deep optical, IR and sub-mm surveys suggest that the peak of 
the cosmic star-formation lies at redshifts on the order of 1-2 or more
\citep{bla99}.  White \& Ghosh (1998, hereafter WG98) predicted that X-ray
binaries that form from this early star-formation activity should result in
the enhancement of X-ray emission in galaxies, most notably around a redshift
of 0.5-1.0 as the massive stars supernova forming low-mass X-ray binaries
(LMXRBs) containing neutron stars. or blackholes  The
LMXRBs containing neutron stars in turn evolve to form millisecond radio
pulsars (MRP), resulting in the observed 
distribution of XRB and MRP in the Galaxy and nearby galaxies.   Further
modeling by Ghosh \& White (2001; hereafter GW01) has been undertaken to
refine their previous 
results, with the primary improvement being the use of more recent
star-formation rate (SFR) history estimates (e.g., Blain et al. 1999).  These
more recent SFR history estimates take into account sub-mm data which is
sensitive to dusty (and hence obscured) high-redshift galaxies that would be
missed by optical/UV surveys.  The main impact of these modifications is that
the current estimate of the enhancement of X-ray flux from galaxies (due to
LMXRB) peaks at $z \sim 1.5$ rather than $z \sim 1.0$.  The modeling of WG98
and GW01 also includes the predictions for enhancements of high-mass X-ray
binaries (HMXRB).  We defer evaluation of the HMXRB enhancement for future
work, but note here that since HMXRB have a much shorter evolutionary time
scale than LMXRB, the HMXRB enhancement (along with other effects of starburst
activity) would closely track the SFR history and therefore occur at higher
redshifts than the LMXRB enhancement discussed here.

A test of these X-ray evolution models would be to evolve the local universe
X-ray luminosity function and derive the number counts of galaxies expected at
given flux levels (i.e., in a logN-logS diagram) for comparison with deep
X-ray surveys.  Unfortunately, the X-ray luminosity function of nearby
galaxies is not known directly because the typical X-ray flux of galaxies is
of order of $\log F_X = -13$ -- $-14 \rm \ ergs \ cm^{-2} \ s^{-1}$ which
is below the limiting flux of existing X-ray all-sky surveys (e.g, the ROSAT
All-Sky Survey has a limiting flux on the order of 
$\log F_{0.5-2.0 \rm keV} \sim -12$ -- $-13 \ \rm  \ ergs \ cm^{-2} \
s^{-1}$; Voges et al. 1999).  Here we take an alternative approach which is to
use the known X-ray/optical luminosity correlation given in \citet{dav92},
determined from Einstein
data, to estimate the X-ray luminosities and fluxes of galaxies in the HDF-N,
where optical luminosities and redshifts have been established.  Since the
expected X-ray flux of high-redshift galaxies is low, \citet{bra01}
performed a stacking analysis of a {\it Chandra} observation of the HDF-N.  We
show 
below that the X-ray flux distribution derived from our analysis is consistent
with the galaxy flux signal detected by {\it Chandra}.  A secondary goal of
this 
paper is to estimate the logN-logS distribution of galaxies (based on the
HDF-N) in order to predict X-ray number counts that could be detected by more
sensitive surveys.

\section{Methodology}
As discussed above, our main goal is to estimate the X-ray flux and luminosity
for galaxies based on their optical luminosity, the X-ray/optical luminosity
correlation and X-ray luminosity evolution models.
The HDF-N ``proper'' sample presented in \citet{coh00} contains redshifts and
R-magnitudes for 125 galaxies spanning $\rm 4.75 \ arcmin^{2}$.
The X-ray/optical correlation given in \citet{dav92} is based on B-band
luminosities, so accordingly the observed R-band magnitudes must be converted
to rest-frame B-band magnitudes.  
We used the k-correction plots in \citet{fre94} which include R to B band
corrections based on optical and UV observations of galaxies 
(which we caution only extends to z=0.6), from which we derived the B-band
luminosity for each galaxy (we used $H_0 = 70 \rm \ km \ s^{-1} \ Mpc^{-1}$
and
$q_0 = 0.1$ throughout this paper). Here we are primarily interested in the
mean properties of galaxies and accordingly did
not segregate by galaxy type, which we differ to future work (see also below).
The k-corrections as a function of galaxy type vary by $\sim 1$ magnitude,
which is somewhat larger than the scatter found for control galaxies by
\cite{fre94}, and accordingly we assume an uncertainty of 1 magnitude.


These B luminosities were then converted to X-ray luminosities using the
$\rm L_{0.5-4.5 \ keV}/L_B$ relation given in \citet{dav92}.
Assuming a power-law spectrum with a photon index of 1.8 and neutral
absorption $N_H < 10^{21} \ \rm cm^{-2}$ (the galactic column towards the
HDF-N is $1.6 \time 10^{20} \rm cm^{-2}$; Stark et al. 1992), which is
consistent with an 
XRB-dominated spectrum and the observed 2-10 keV X-ray spectra of nearby
galaxies (c.f., Ptak et al. 1999 and references therein), $F_{2-10 \rm \
keV} = F_{0.5-4.5 \rm \ keV}$ to within $\sim 
15\%$. Similarly, these same spectral assumptions also imply a $F_{0.5-2.0 
\rm keV}$ flux that is $\sim (0.5-0.6)F_{0.5-4.5 \ \rm keV}$ (at redshifts
exceeding 0.5, $F_{0.5-2.0 \ \rm keV} = 0.55 F_{0.5-4.5 \ \rm keV}$ for all
$N_H < 10^{21} \ \rm cm^{-2}$) and we adopt 1.0 and 0.55 as the conversion
factor to
the 2.0-10.0 keV and 0.5-2.0 keV bandpasses.  We also note
that the stellar evolution models we are concerned with here are relevant only
on long look-back time scales, and therefore only the lower-mass, older stellar
populations have a significant impact in our analysis.   
Ignoring the younger stellar populations
biases our results in the sense of underestimating the expected X-ray flux
for starburst systems which tend to exhibit enhanced
supernovae and high-mass X-ray binary populations (see Fabbiano 1989).
Since the \citet{dav92} correlation was performed using all
galaxies types, the differing amount of X-ray production as a function of
galaxy type inherently contributes to the scatter of the correlation.
Accordingly, the results of this paper should be accurate in a statistical
sense but should not be applied to any individual galaxy.  

The resultant B and X-ray luminosity distributions
derived from this procedure are shown in Figure 1.  The errors given in
this figure are derived from a Monte-Carlo approach in which the conversion
to the B and X-ray bandpasses was repeated (1000 iterations) with gaussian
deviates with a standard deviation of 1 magnitude added to the resulant B
magnitudes and gaussian deviates with standard deviation of 1 added to the log
X-ray luminosities (i.e., simulating an order-of-magnitude scatter in
$L_X/L_B$).   Errors derived in this way were comparable to the counting
errors that would be expected in each histogram bin (i.e., $\sqrt{N}$,
where N = the number of galaxies in a given histogram bin).  
The mean galaxy 
$L_{2-10 \rm \ keV}$ is $3.8 \times 10^{39}$ \ergs, which is comparable to the mean value for
passive galaxies given in \citet{geo99} of
$3.2 \times 10^{39}$ \ergs (after adjusting to our values of $H_0$ and $q_0$).
These values are somewhat lower than other nearby galaxies X-ray luminosity
estimates (e.g., Fabbiano, Trinchieri, \& McDonal 1984) however it should
noted that here and in \citet{geo99} active (including narrow-line) and
starburst galaxies have not been (explicitly) included in the analyses.
\begin{figure*}[htbn]
\epsscale{1.5}
\plottwo{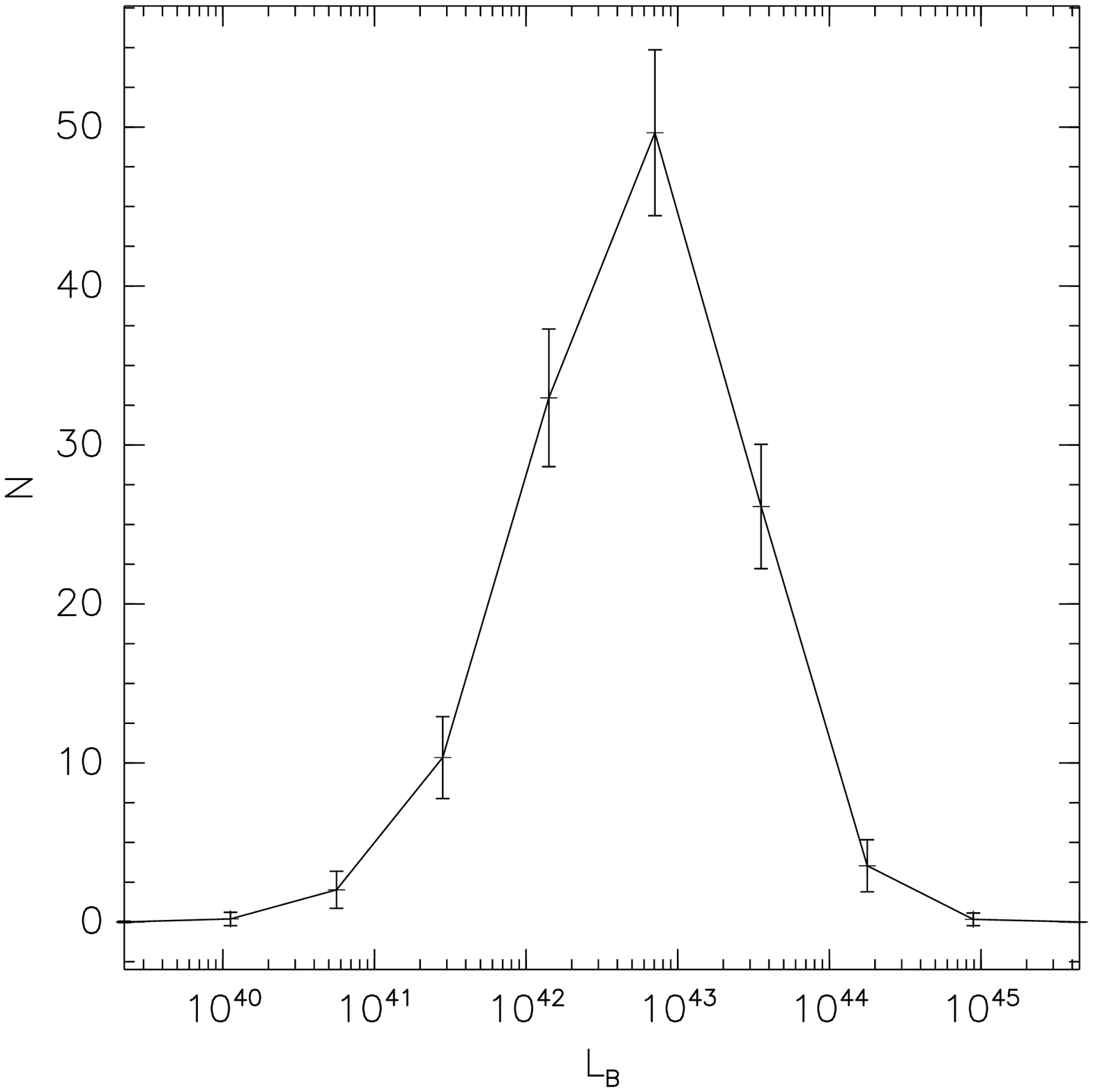}{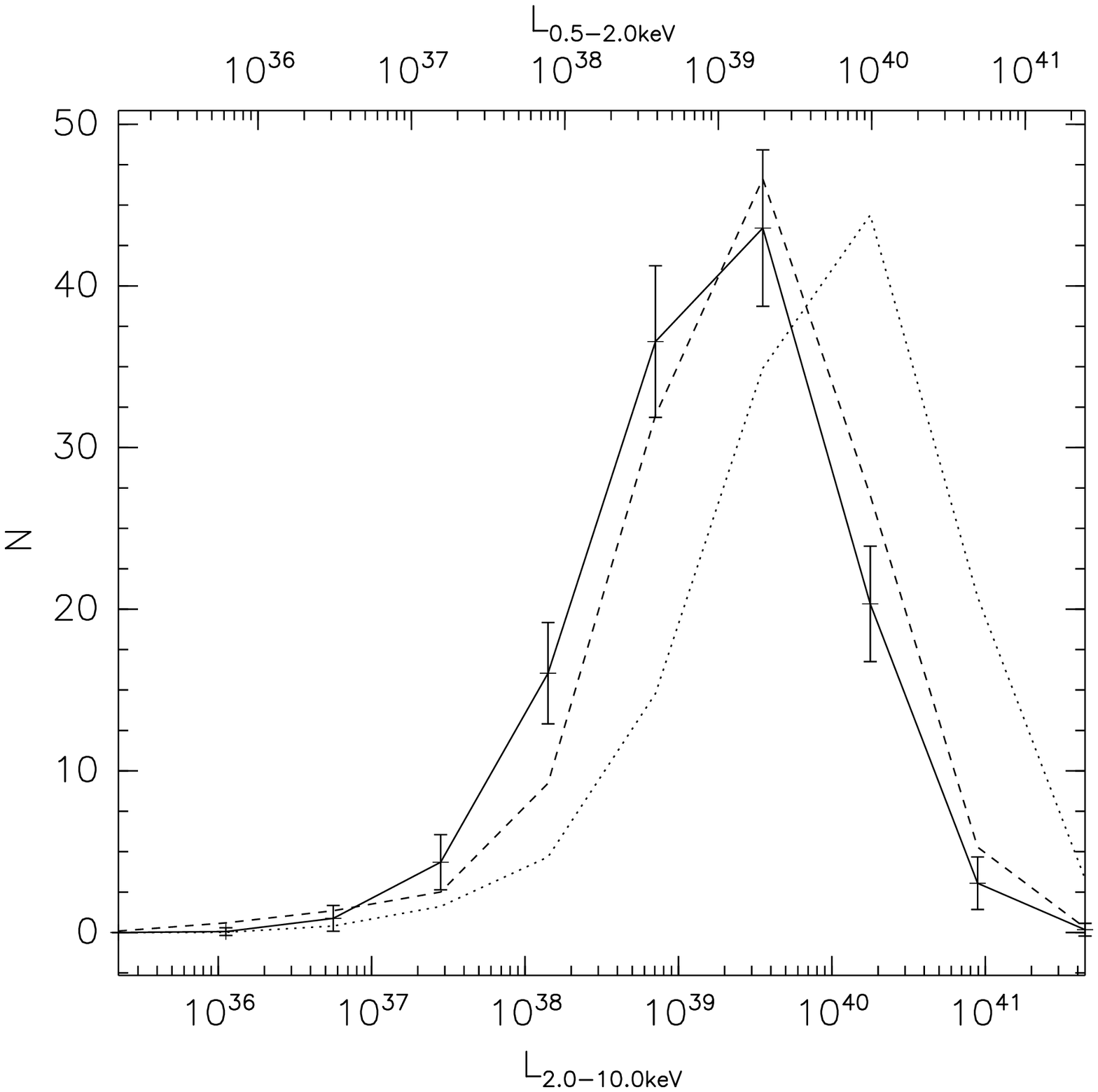}
\caption{The estimated B-band (left) and X-ray luminosities (right)
for the HDF-N galaxies.  The error bars shown are derived from simulations
that incorporate a 1 magnitude uncertainty in the k-corrections and an order of magnitude uncertainty in the X-ray/optical flux ratio.  
The X-ray luminosity distribution is given for the unevolved case (solid line
and points) as well as that derived from the ``Peak-M'' (dashed line) and
``Gaussian'' (dotted line) evolution models.}
\end{figure*}


\subsection{Evolution in $L_X/L_B$}
We proceed now assuming that the ratio $\rm L_X/L_B$ evolves as a result of an
excess of X-ray binaries at earlier epochs.  
We use the evolutionary models ``Peak-M'' and ``Gaussian'' given in
\citet{gho01}, 
plotted in Figure 2.  Both models are based on fits to the observed
star-formation rate using optical/UV observations, however the ``Gaussian''
model also includes a
component with a Gaussian functional form that takes into account the
evolution of the IR luminosity 
function using IR and sub-mm data (c.f., Blain et al. 1999).
These models only take into account the relative change
in X-ray luminosity as a function of redshift, however in this paper we are
attempting to derive the X-ray luminosities of galaxies based on the
rest-frame B-band 
luminosity.  Accordingly, any optical evolution must be taken into
account explicitly before we could apply our X-ray evolution models.  
To this end, we took the mean and standard deviation of the
B-band luminosities in redshift bins large enough to contain at least 20
galaxies and fit $\log L_B$ as a function of z with a linear model
(also shown in Figure 2).  The X-ray luminosities were evolved by
multiplying them by the ratio $E_X(z)/E_B(z)$, where $E_X(z)$ and $E_B(z)$ are
the X-ray and optical (after normalizing to z=0) evolution models shown in
Figure 2.  The net amount of optical evolution derived in this way amounts to
only a factor of $\sim 2$ from a redshift of 0 to 1.
\begin{figure*}[htbn]
\plottwo{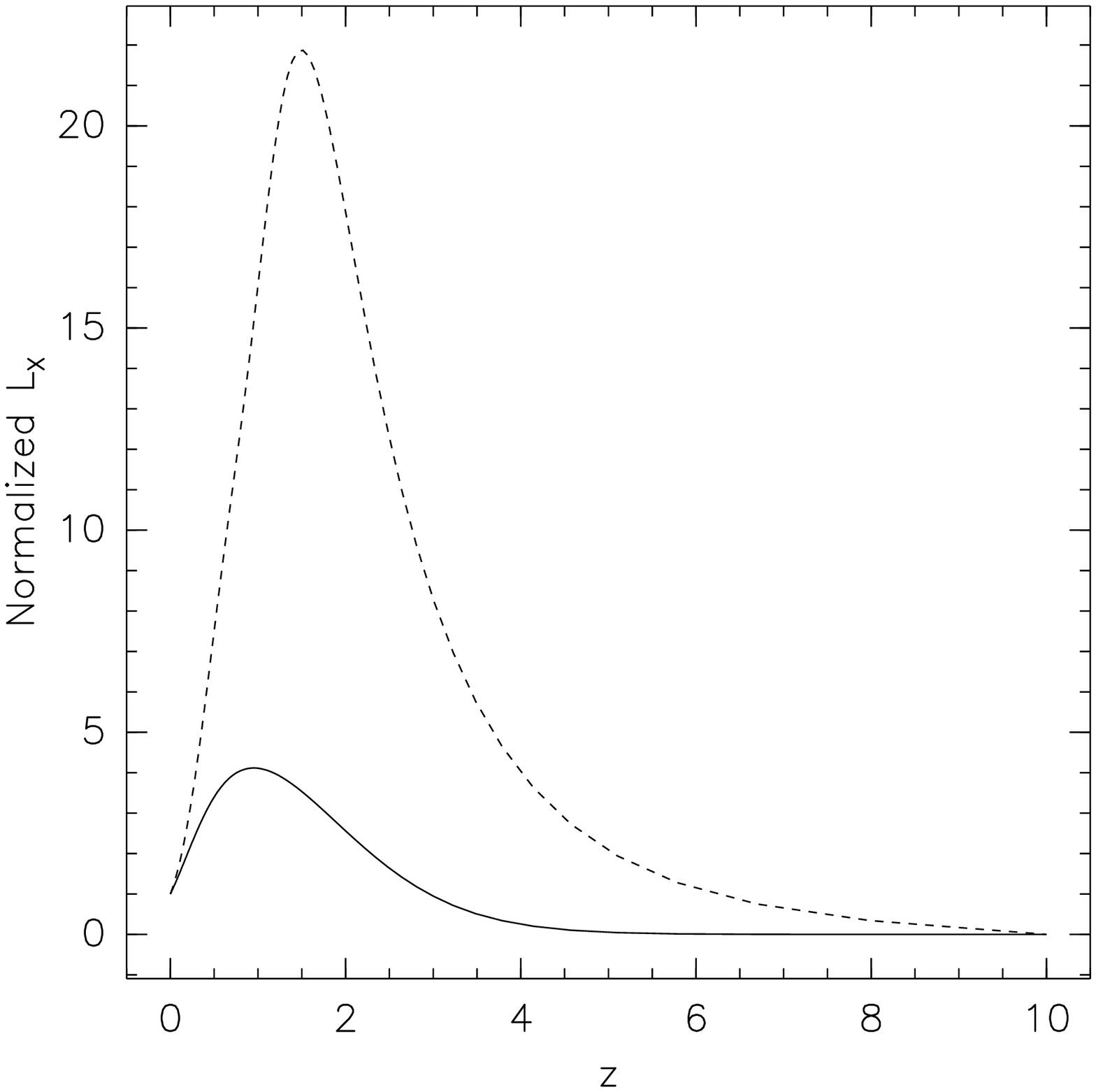}{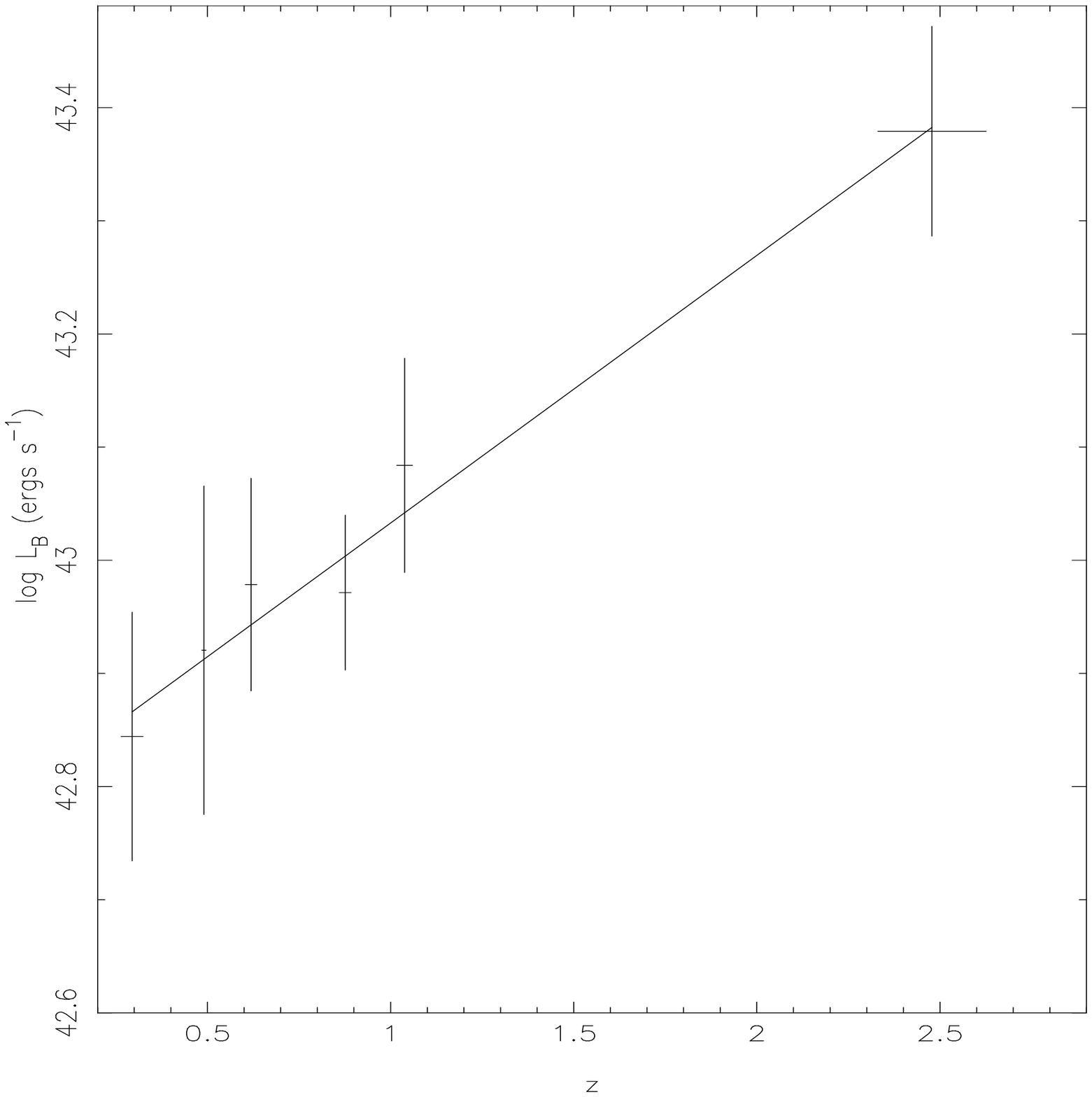}
\caption{(left) Evolution models ``Peak-M'' (solid line) and
``Gaussian'' (dashed line) for the 
X-ray luminosity of normal galaxies, normalized at z=0, from
\citet{gho01}. (right) Observed evolution in mean B-band luminosity
for the HDF-N galaxies, with the linear fit to $\log L_B$ versus z plotted
as a solid line.  The error bars shown were derived by the dispersion of
luminosities in each redshift bin.}
\end{figure*}

\section{Results}
Applying these procedures, the mean
2-10 keV X-ray galaxy flux increased from $2.3 \times 10^{-18}$ \ergcms~ to 4.2
and $8.4 \times 10^{-18}$ 
\ergcms~ and the mean 2-10 keV X-ray luminosity increased from $3.8 \times
10^{39}$ \ergs~ to 5.3 and $19. \times 10^{39}$ \ergs~ for the ``Peak-M'' and
``Gaussian''
models, respectively (with 0.5-2.0 keV values a factor of 0.55 lower as
discussed above).  The distribution of $L_X$ and $F_X$ (before and after 
including evolution) is shown in Figures 1 and 3.  The flux density implied is
4.3 and $7.8 \times 10^{-13} \ \rm ergs \ s^{-1} \ cm^{-2} \ deg.^{-2}$ in the
0.5-2.0 keV and 2-10 keV bandpasses, or 2-4\% of the 2-10 keV X-ray
background.  \citet{kun01} find that the extra-galactic X-ray backround in the
1-2 keV bandpass is $\sim 5.7 \times 10^{-12} \ \rm ergs \ s^{-1} \ cm^{-2} \
deg.^{-2}$.  With same spectral assumptions discussed above, $F_{1-2 \rm
\ keV} \sim 0.3F_{0.5-4.5 \ \rm keV}$, implying a 1-2 keV flux density of $\sim
2.3  \times 10^{-13} \ \rm ergs \ s^{-1} \ cm^{-2} \ deg.^{-2}$, or $\sim
4\%$ of the observed 1-2 keV background.
We repeated our analysis after only including
galaxies with redshifts in the range of 0.5-1.0 (55 galaxies) which 
resulted in a mean 2-10 keV X-ray
flux of $\sim 7.0 \times 10^{-18}$  \ergcms (indicating that our results are
dominated by galaxies in this redshift range).  
\begin{figure*}[htbn]
\epsscale{0.8}
\plotone{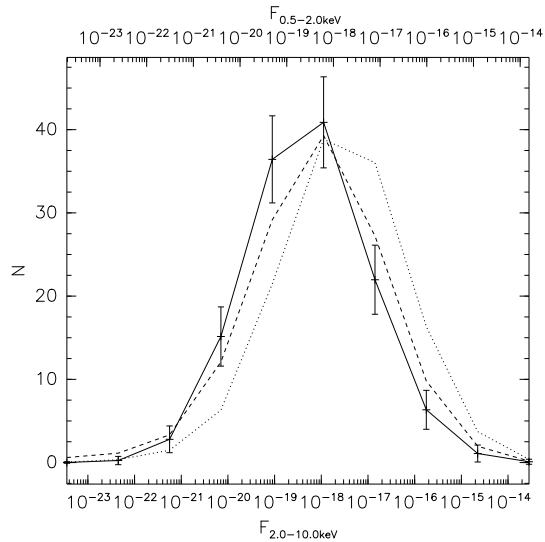}
\caption{0.5-2.0 keV and 2-10 keV flux distributions predicted for 
HDF-N galaxies.   The lines and errors are defined as in Figure 1.}
\end{figure*}

We proceed now to the question of the numbers of galaxies expected
to be detected as a function of flux in ultra-deep {\it Chandra}, 
{\it XMM-Newton}, or
future X-ray surveys.  To address this, we derived the ``logN-logS''
number count distribution (i.e., the number of galaxies exceeding flux S as a
function of flux)
implied by our analysis, shown in Figure 4, using the HDF-N area sited above. 
The galaxy number counts have been corrected for spectroscopic completeness
using Figure 1 from \citet{coh00}.
This figure also shows the 
fits to the 0.5-2.0 keV and 2-10 keV HDF-S logN-logS distrubitions
\citep{toz01} (note that these curves are extrapolations below fluxes of $\sim
10^{-16}$ and $10^{-15} \rm \ ergs \ cm^{-2} \ s^{-1}$ for the soft and hard
bands, respectively).
The 2-10 keV HDF-S logN-logS in the $S=10^{-15} - 5 \times
10^{-14}$ \ergcms~ 
range is in good agreement with the HDF-N logN-logS curve given in
\citet{gar01}, where fluctuation analysis implies that the logN-logS curve 
flattens somewhat below $S=1.0 \times 10^{-15}$ \ergcms.  The total (i.e.,
including AGN) source densities determined by \citet{bra01} are marked
which also implies a flattening of the X-ray logN-logS.  Note that the results
of the stacking analysis in \citet{bra01} cannot be plotted on this
diagram since by design the analysis was done around optically-detected
galaxies without significant X-ray counterparts and hence the area is
not well defined.
Depending on the amount that the AGN logN-logS flattens, the total X-ray 
logN-logS may become dominated by normal galaxies at fluxes below 
$\log F_X = -17$ -- $-18$ \ergcms, particularly in the soft band.  
\begin{figure*}
\epsscale{2.0}
\plottwo{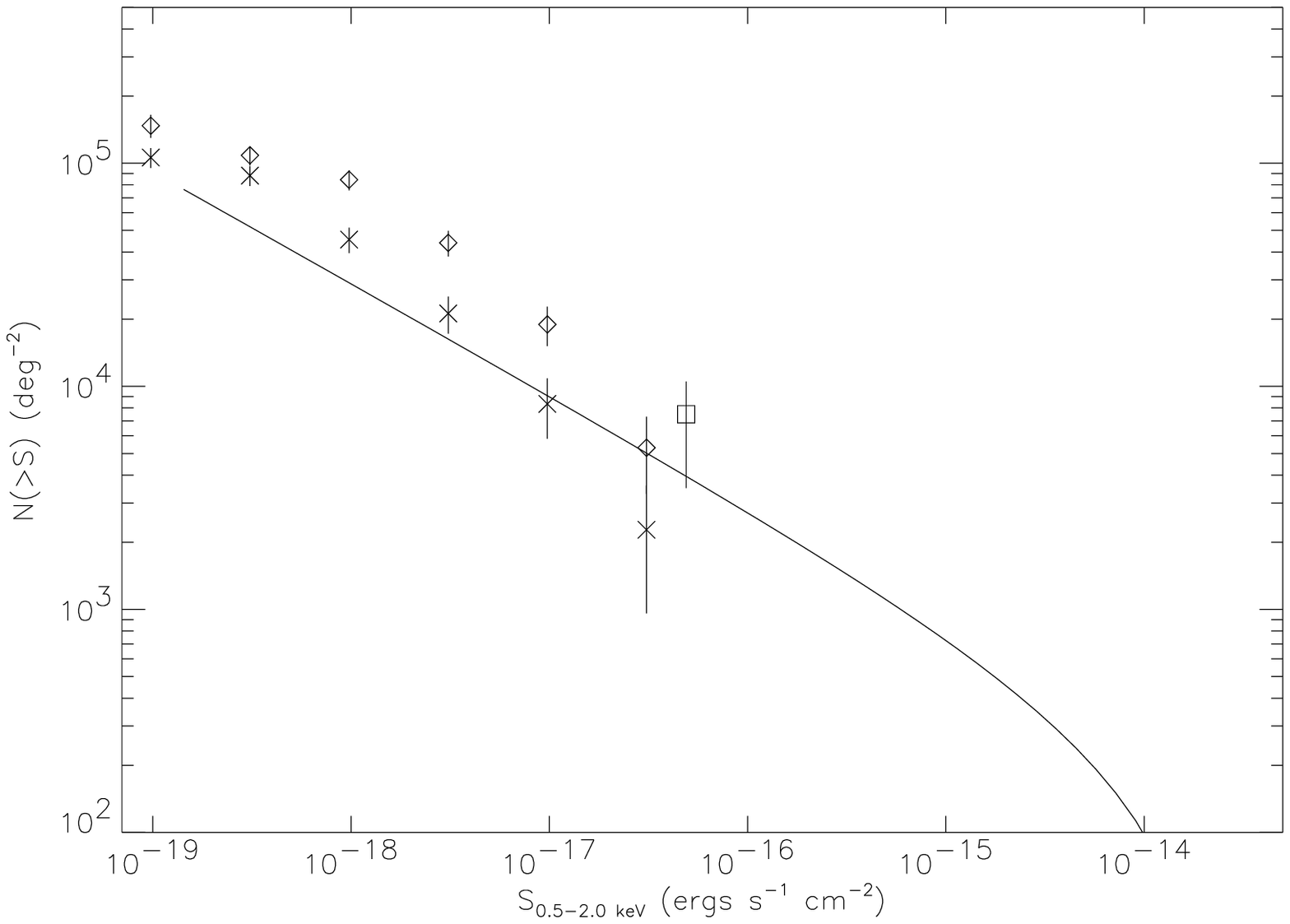}{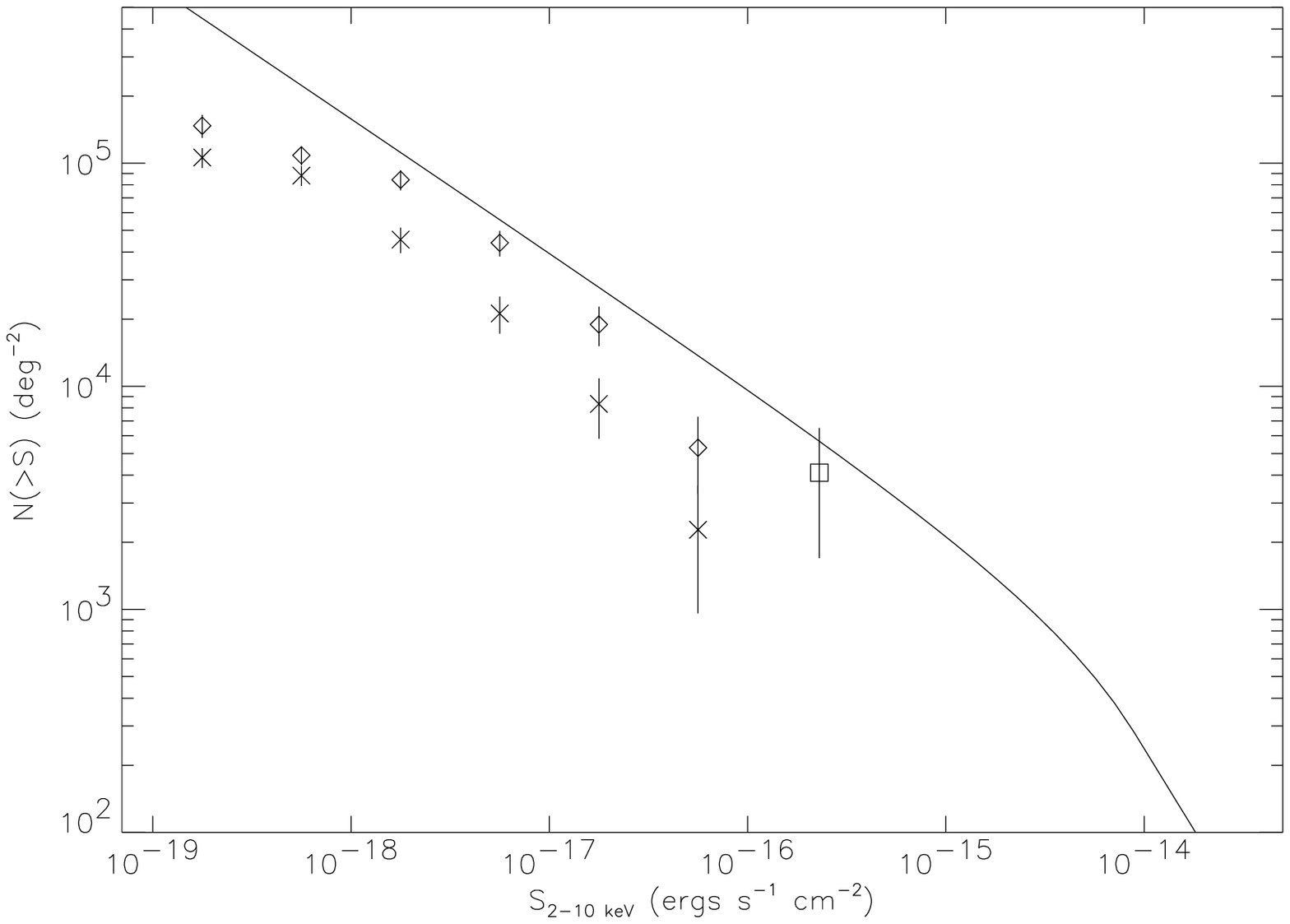}
\caption{The logN-logS for galaxies derived from the HDF-N estimates for the
``Gaussian'' (diamonds) and ``Peak-M'' evolution models in the 0.5-2.0 kev
(left) and 2.0-10.0 keV (right) bandpasses.  The solid lines give the
logN-logS double power-law fit
from \citet{toz01} based on the Chandra observation of the HDF-S (note that
this logN-logS includes all contributions to number counts).   
The square points mark the source density determined
in the 500 ks Chandra HDF-N observation \citep{bra01}. $F_X = 10^{18}$
\ergcms~ corresponds to $R\sim24$.}
\end{figure*}

\section{Discussion}
We have estimated the X-ray fluxes of galaxies in the Hubble Deep Field using
the known X-ray/B-band luminosity correlation and taking evolution in the
X-ray binary population of the galaxies into account (using the ``Gaussian''
and ``Peak-M'' models in GW01).  Here we are only attempting to estimate the
LMXRB contribution to the galaxy fluxes, which should dominate the X-ray
fluxes unless significant starburst and/or AGN activity is present.
The mean 2-10 keV X-ray
flux that the Gaussian model predicts is $\sim 8.2 \times 10^{-18}$  \ergcms~
which can 
be compared with $\sim 1.7 \times 10^{-17}$ \ergcms ~obtained from a stacking
analysis of a 500 ks {\it Chandra} observation of the HDF-N (converted to the
2-10 keV bandpass from a signal of $2.3 \times 10^{-17}$ \ergcms~ in the
0.5-8.0 keV bandpass assuming the same spectrum as discussed above) determined
by \citet{bra01}.  The ``Peak-M'' model predicts a mean X-ray flux a factor of
$\sim 3.5$ lower than the ``Gaussian'' model.
The ``Gaussian'' model is therefore predicting a signal which is a factor of
$\sim 2$ less than the signal observed, although we note that \citet{bra01}
selected optically-bright galaxies (a sample of 11) for their stacking
analysis, and our estimate of the mean galaxy X-ray luminosity 
(prior to the application of any
evolution) is evidently somewhat lower than that assumed by \citet{bra01} but
is nevertheless consistent with other estimates.  In both the hard and soft
bandpass, we are predicting that galaxies make up $\sim 4\%$ of the X-ray
background down to a flux of $\sim 10^{-18} \ \rm ergs \  cm^{-2} \ s^{-1}$.

The best test of these models would be the direct detection of high-z galaxies
in the X-ray bandpass, obviously with sufficient spectroscopic 
identification to
rule out an AGN contribution to the X-ray flux.  
This would require an X-ray detection sensitivity on the order of $10^{-17}$
\ergcms, which would imply an exposure on the order of 10 Ms for 
{\it Chandra}
(i.e., the effective exposure in the stacking analysis was 5 Ms), while even
such a large exposure would be insufficient for {\it XMM-Newton} to reach these
detection sensitivites (due to {\it XMM-Newton} becoming background-limited at
very large exposures).  Optical (and, to a lesser extent, IR) spectroscopic 
identifications
could possibly miss highly-obscured AGN or low-luminosity AGN which would
produce enhanced 
X-ray fluxes relative to normal galaxies. A simple expectation is that the
X-ray emission of galaxies would be extended (on spatial scales comparable to
the optical emission), while the X-ray image of galaxies dominated by AGN
would of course be unresolved.  The half-light radius 
of galaxies in the HST Medium Deep Survey was typically $\sim 0.1-1.0''$,
with a median value around $0.6''$ \citep{rat99}.  Arcsecond resolution will
therefore be necessary to detect any extension to galaxies.
\acknowledgements
We thank the referee, Paulo Tozzi for useful comments that strengthened
this paper.  This research was
partly funded by NASA contract NAS8-38252 to Pennsylvania State University.

\end{document}